%% file: main.tex
\documentclass[acmsmall,screen]{acmart}
\AtBeginDocument{%
  }
\usepackage{algorithmic}
\usepackage{color}
\usepackage{multirow}
\usepackage[linesnumbered,algoruled,nofillcomment]{algorithm2e}
\usepackage{subfigure}
\usepackage{colortbl}
\definecolor{xgray}{gray}{.9}
\newcommand{\toolname}{\textsf{DRIVE}}

\def\code#1{\texttt{#1}}

\setcopyright{acmcopyright}
\copyrightyear{XXXX}
\acmYear{XXXX}
\acmDOI{}
\acmJournal{TOSEM}
\acmVolume{0}
\acmNumber{0}
\acmArticle{0}
\acmMonth{0}

\begin{document}

\title{\toolname: Dockerfile Rule Mining  and Violation Detection
}
\author{Yu Zhou}
\email{zhouyu@nuaa.edu.cn}
\author{Weilin Zhan}
\email{zhanweilin@nuaa.edu.cn}
\author{Zi Li}
\email{zl.2021@nuaa.edu.cn}
\affiliation{%
	\institution{Nanjing University of Aeronautics and Astronautics}
	\city{Nanjing}
	\country{China}
}

\author{Tingting Han}
\email{t.han@bbk.ac.uk}

\author{Taolue Chen}
\email{t.chen@bbk.ac.uk}
\authornote{Corresponding author.}
\affiliation{%
	\institution{Birkbeck, University of London}
	\city{London}
	\country{UK}}

\author{Harald Gall}
\email{gall@ifi.uzh.ch}
\affiliation{%
	\institution{University of Zurich}
	\city{Zurich}
	\country{Switzerland}
}

\begin{abstract}
	A Dockerfile defines a set of instructions to build Docker images, which can then be instantiated to support containerized applications. Recent studies have revealed a considerable amount of quality issues with Dockerfiles. In this paper, we propose a novel approach {\toolname} (\textbf{D}ockerfiles \textbf{R}ule m\textbf{I}ning and \textbf{V}iolation d\textbf{E}tection) to mine implicit rules and detect potential violations of such rules in Dockerfiles. {\toolname} firstly parses Dockerfiles and transforms them to an intermediate representation.  It then leverages an efficient sequential pattern mining algorithm to extract potential patterns. With heuristic-based reduction and moderate human intervention, potential rules are identified, which can then be utilized to detect potential violations of Dockerfiles.  {\toolname} identifies  34 semantic rules and 19 syntactic rules including 9 new semantic rules which have not been reported elsewhere. 	Extensive experiments on real-world Dockerfiles demonstrate the efficacy of our approach.
\end{abstract}

\maketitle


\input{sections/intro}
\input{sections/bg}

\input{sections/approach}

\input{sections/exp}
\input{sections/discuss}
\input{sections/related}

\section{Conclusion}\label{conclusion}
In this paper, we present {\toolname}, a novel approach to efficiently mine implicit rules from high-quality Dockerfiles, based on sequential pattern mining techniques. We demonstrate the efficacy of our approach against state-of-the-art baselines. {\toolname} can find more useful implicit rules with less time, among which 9 rules have been firstly reported. Since Dockerfiles can also be reused by some other Docker-compatible containers (e.g., Podman), our approach also has potentials to improve the quality of built images of those containers.

In the future, we plan to augment {\toolname} with more functionalities, such as repair recommendations for detected violations, and develop full-fledged tools (as plugin of mainstream IDEs) to deliver better usability. More generally, we believe that such a data-driven paradigm could be also applied to other related areas, such as configuration pattern mining and code smell detection.

\begin{acks}
This work was partially supported by the National Natural Science Foundation of China (NSFC, No.\ 61972197), the Natural Science Foundation of Jiangsu Province (No.\ BK20201292), the Fundamental Research Funds for the Central Universities (No.\ NG2023005), and the Collaborative Innovation Center of Novel Software Technology and Industrialization. T. Chen is partially supported by Birkbeck BEI School Project (EFFECT), an oversea grant from the State Key Laboratory of Novel Software Technology, Nanjing University (KFKT2022A03),  National Natural Science Foundation of China (Grant No.\ 62272397).
\end{acks}

\bibliographystyle{ACM-Reference-Format}
\bibliography{draft}

\end{document}

%% file: sections/intro.tex
\section{Introduction}\label{intro}
Virtualization plays a fundamental role in 
cloud computing~\cite{bernstein2014}. Comparing with traditional virtualization techniques (e.g., hypervisor), containerization is a light-weight and efficient alternative, gaining increasing popularity in practice~\cite{morabito2015,pahl2015}. Nowadays, Docker has become a mainstream supporting tool for containerization of applications. According to a recent report~\cite{dockerindex}, as of August 31, 2021 there has been a total of 396 billion all-time pulls on Docker Hub, up from 318 billion just six months ago, an increase of about 25\% year-over-year. 

Docker relies on \emph{Docker images} to deliver deployable applications. Since the corresponding execution environment is also encapsulated in the images, users could run the applications on target platforms directly without considering configuration differences.
The instructions of building Docker images are specified in order in \emph{Dockerfiles} according to a set of syntax rules. As a result, the quality of Dockerfiles is crucial to the success of built images. However, recent empirical studies on large-scale open-source projects have exposed serious concerns on the quality of existing Dockerfiles in relation to either their functionality or performance, some of which are even broken~\cite{henkel2020, henkel2021,ksontini21}. 

Clearly, Dockerfiles, like other source-level artifacts, need to be carefully designed following basic principles, rules, or otherwise patterns in a practical term. Several tools, such as VSCode plugins,\footnote{\url{https://code.visualstudio.com/docs/containers/overview}} provide preliminary support for Dockerfile construction, but remain at the syntax level (e.g., highlighting keywords, hovering tips, etc.). Indeed, the official Docker website provides practice guidelines for writing Dockerfiles~\cite{bestpractice}. However, such guidelines are at a high level and of general-purpose, and, most importantly, focus on Docker-specific commands only. In Dockerfiles, Shell commands (i.e., those led by the \code{RUN} command) are most frequently used, which usually account for over 40\% of all the instructions (with some empirical study even reveals that up to 68.3\% of Dockerfile changes focuses
on the Shell commands~\cite{wu2020dockerfile}) and about  90\% of repositories use Shell commands~\cite{cito2017}. 

\begin{figure}[h]
 	\centering
 	\vspace*{-3mm}
 	\includegraphics[scale=0.65]{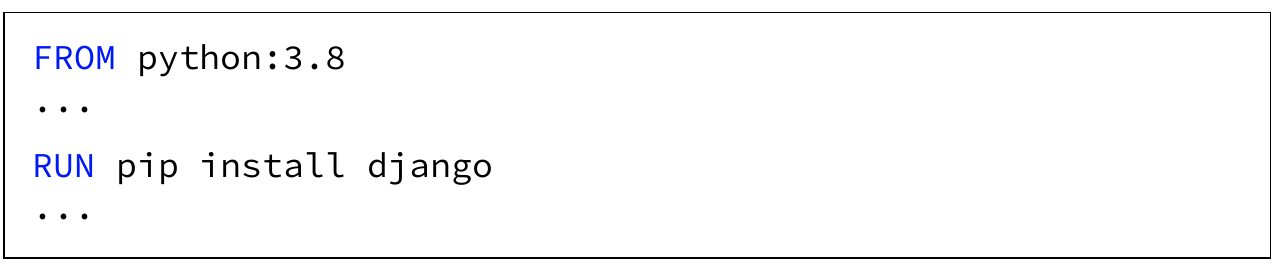}
 	\caption{Pip without --no-cache-dir argument}
 	\vspace{-3mm}
 	\label{exam1a}
 \end{figure}

\begin{figure}[h]
 	\centering
 	\includegraphics[scale=0.65]{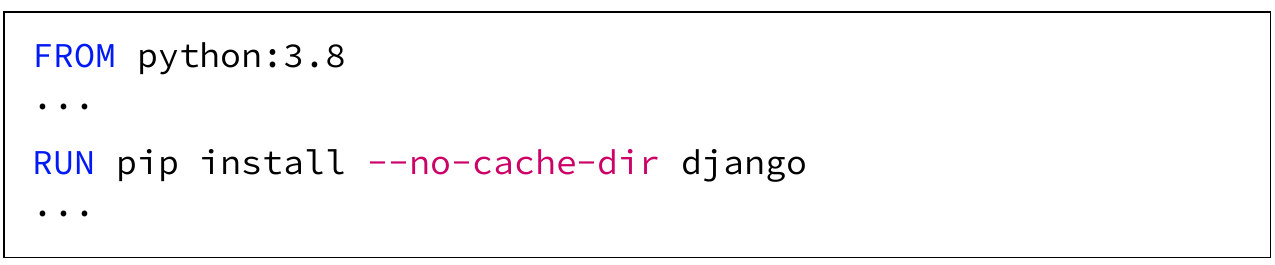}
 		\vspace{-3mm}
 	\caption{Pip with --no-cache-dir argument}
 	\label{exam1b}
 \end{figure}

\begin{figure}[h]
 	\centering
 	\vspace*{-3mm}
 	\includegraphics[scale=0.66]{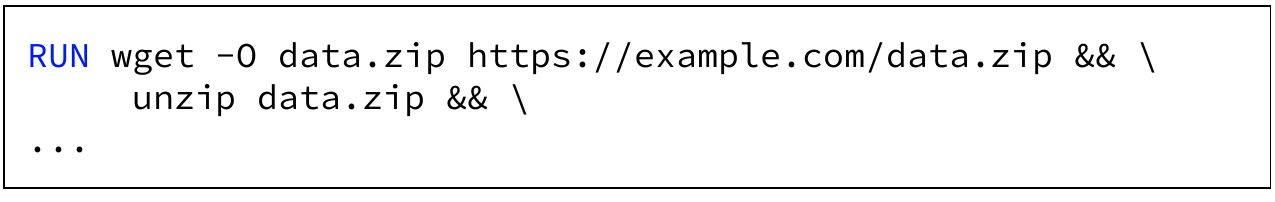}
 	\caption{Unzip w/o remove instruction\label{exam2a}}
 	\vspace{-4mm}
 	
 \end{figure}

Fig.~\ref{exam1a} and Fig.~\ref{exam2a} give two concrete examples taken from real-world Docker projects. 
In Fig.~\ref{exam1a}, the required software dependency is installed through the \code{pip install} command in a regular Python containerization project. Normally, 
\code{pip install [packages]} would suffice in most cases of traditional environments. However, in Docker it will cause  performance issues, although it can pass the syntax check and be built successfully. The resulting Docker image would be larger than necessary, which is caused by the default caching mechanism provided by the \code{pip} command to reduce the amount of time spent on duplicated downloads and builds. This mechanism has unexpected side-effects in a Docker image, since the image is usually built once and Docker itself provides a separate caching mechanism. To save space, we can add  the \code{--no-cache-dir} flag (Fig.~\ref{exam1b}). 

%

\begin{figure}[h]
 	\centering
 	\vspace*{-3mm}
 	\includegraphics[scale=0.66]{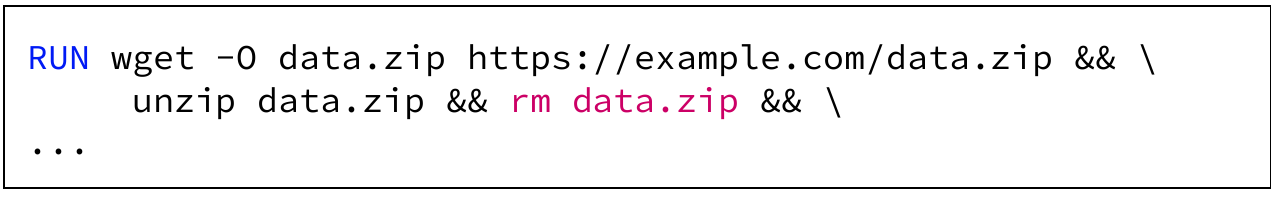}
 	\caption{Unzip with remove instruction\label{exam2b}}
 	
 \end{figure}

Fig.~\ref{exam2a} shows a more sophisticated example with multiple commands. Frequently, we need to download compressed files from the Internet followed by an uncompressing command. In Fig.~\ref{exam2a}, \code{wget} is used to retrieve a zip file in a remote website, and then \code{unzip} is used to uncompress the downloaded file.  Executing such a command will retain the original zip file and create a new folder to place the extracted files, but the original zip file is still kept which is no longer needed. Such inadvertent inclusion of unnecessary files in images inevitably results in longer build time and larger image size. Therefore, the original downloaded file should be deleted afterwards with an additional \code{rm} command as illustrated in Fig.~\ref{exam2b}. 

%

The above examples demonstrate that there are some implicit rules which should be respected when writing Dockerfiles. Unfortunately, these rules are largely ignored by the official best practice guidelines, and are frequently violated even in  some high-rated real-world projects. For example, in the Dockerfile of a top-rated face alignment project,\footnote{\url{https://github.com/1adrianb/face-alignment/blob/master/Dockerfile}} the authors do not clean the cache after using \code{conda install} (cf.\ Rule 25 in Table~\ref{tab:rules}). Similar violation of this rule can be found in the Dockerfile of another popular project.\footnote{\url{https://github.com/zjhuang22/maskscoring_rcnn/blob/master/docker/Dockerfile}} The violation of such rules may not necessarily lead to build failures, but may have a negative effect on non-functional properties instead, which is similar to the notion of ``code smells" in programs~\cite{rasool2015review}. 

Some work and tools have been proposed to address this issue. The two representative tools are \emph{Hadolint}~\cite{hadolint} and \emph{Binnacle}~\cite{henkel2020}, which attempt to identify patterns in Dockerfile commands. However, they both suffer from various limitations such as heavy human intervention and low efficiency. For example, in \emph{Hadolint}, the patterns are mainly  specified by community experts without automatic pattern discovery mechanism. Moreover, these patterns (or rules) which are either Dockerfile-specific or Shell commands are mostly at the syntax level. 
In \emph{Binnacle}, a multi-stage parsing technique, i.e., phased parsing, is utilized to parse Dockerfiles based on abstract syntax trees (ASTs), but the rule mining process still depends on prior knowledge to select sub-trees. Moreover, a severe limitation of this approach is that they can only extract \emph{local} Tree Association Rule (TAR) (i.e., the localness problem), since finding arbitrary TARs is computationally infeasible~\cite{henkel2020}. As a concrete example, the ``remove after downloading" rule in Fig.~\ref{exam2b} cannot be discovered by \emph{Binnacle}, if the related commands are not located in the same subtree (i.e., with the same manually selected root node).



In this paper, we propose a novel approach {\toolname} (\textbf{D}ockerfiles \textbf{R}ule m\textbf{I}ning and \textbf{V}iolation d\textbf{E}tection) to identify general patterns in Dockerfiles with moderate human participation. 
Our approach adopts a sequential pattern mining method.
In particular, it transforms Dockerfiles into intermediate representations on which standard sequential pattern mining algorithm can be applied. This approach can scale up to identify arbitrarily frequent patterns and requires less time compared with the baseline work. As a result,
{\toolname} is able to identify new rules which have not been discovered by previous approaches. Specifically, we produce 9 new rules, and reproduce 19 syntactic and 25 semantic rules which were human summarized before. These implicit rules can serve as specific guidelines for writing Dockerfiles in practice, the usefulness of which is indeed witnessed by, e.g., comments from StackOverflow posts (cf.\ Section~\ref{exp}).  
%
Moreover, given the identified rules, {\toolname} can detect violations of such patterns by analysing input Dockerfiles.

It is worth emphasizing that, our contribution lies in not only the new identified rules, but also the method leading to these findings. Previous work requires Dockerfile experts to summarize the rules which are laborious and time consuming, and, perhaps more importantly, lacks extensibility. These rules were presented via ASTs, which are harder to mine. Our approach largely automates this process, and crucially, is sequence-based (i.e., we treat Dockerfiles as sequences and the mined rules are also formulated as properties of sequences). It can be envisaged that, in the future, more Dockerfiles will emerge, and our approach can be easily applied to produce more useful rules. Such a data-driven nature turns out be indispensable for modern software engineering practice.


In summary, we make the following contributions.

\begin{itemize}
	\item We propose an efficient pattern mining approach for Dockerfiles with moderate human intervention. 
	
	\item We obtain 19 syntactic and 34 semantic rules to encode state-of-art Dockerfile best practice, including 9 semantic rules which have not been reported elsewhere. 
		
		
	\item We present new violation detection algorithms and tool support for  Dockerfiles. 
	
	\item We collect and construct high-quality Dockerfile dataset, which is more diverse and three times larger than the existing one, and  is potentially beneficial for future research. 
\end{itemize}

\noindent\emph{Organization.} The rest of the paper is structured as follows. Section~\ref{bg} briefly introduces the background. Section~\ref{approach} describes our approach in detail. Section~\ref{exp} presents the experimental settings and results. Section~\ref{discuss} discusses the findings further and threats to validity. Section \ref{related} reviews the related work. Section~\ref{conclusion} concludes the paper and outlines future research.

The implementation of our approach, as well as the dataset, is publicly available at \url{https://github.com/zwlin98/DRIVE}.

%% file: sections/bg.tex
\section{Background}\label{bg}
\subsection{Containerization and Docker}
Different from traditional heavy-weight virtualization techniques such as virtual machines (VMs), container-based virtualization, a.k.a.\ containerization, encapsulates specific application files, dependent libraries, runtime support and environmental variables into a separate deployable file system, usually known as an  image~\cite{anderson2015}. Such encapsulation could hide the underlying heterogeneity of the running applications, which can greatly facilitate the practice of infrastructure-as-code (IaC)~\cite{artac2017devops}. Containerization allows for running applications in an isolated environment as an independent process. Multiple processes can share the same operating system (OS) kernel and run simultaneously. Since containerization only includes necessary files to deploy the application, and does not require a complete guest OS copy, this leads to a much reduced file size and greatly enhanced performance. 

Among the many containerization-enabling techniques, Docker is the most popular and \emph{de facto} industry standard nowadays. Dockerfiles direct the building process of Docker images and adopt  a layered construction strategy. Namely, the instructions in Dockerfiles are executed sequentially where the execution of each line generates a branch (or a directory) in the instance's overlay file system, and each corresponds to a layer in the target image~\cite{huang2019}.

\subsection{Sequential pattern mining} \label{sect:spm}
Pattern mining aims to find interesting patterns in a dataset. Various mining techniques have been proposed in the literature, such as frequent itemset mining~\cite{luna2019} and association rule learning~\cite{slimani2014}. 
{\toolname} mainly adopts a sequential pattern mining approach~\cite{fournier2017} in which the order information of items is preserved. Generally, sequential pattern mining aims to identify frequent subsequences out of a sequence dataset. A frequent subsequence $s$ is usually defined as a subsequence whose support value $support(s)$ exceeds a pre-defined threshold $t$.  Exhaustive enumeration of all subsequences would be practically infeasible. 
{\toolname} utilizes an efficient  algorithm, i.e., PrefixSpan, to identify sequential patterns~\cite{pei2004}. Different from typical Apriori like methods~\cite{srikant1996}, the basic idea of PrefixSpan is to examine only the prefix subsequences and project only their corresponding suffix subsequences into projected databases. It explores two kinds of database projections to improve the efficiency and an additional main-memory-based technique is developed to further speed up the performance~\cite{pei2004}. PrefixSpan represents one of the fastest sequence mining algorithms, and is widely used in practice.

\subsection{Dockerfile linters}
\emph{Hadolint} is perhaps the most popular open-source Dockerfile linter currently~\cite{hadolint}. It employs static analysis  techniques to  identify and fix issues in Dockerfiles, improving the quality and security of Docker images. \emph{Hadolint} includes a rich set of rules that can be customized according to user needs. These rules are derived from Dockerfile best practices and expert experience, which can be regarded as domain knowledge. \emph{Hadolint} leverages both the Dockerfile parser and the Shell parser to implement specific detection methods for the violation of each rule.

\emph{Binnacle} is a another tool which can be used to excavate and enforce Dockerfile rules~\cite{henkel2020}. To this end, it first builds an abstract syntax tree for the Dockerfile using multi-stage parsing techniques, selects nodes of interest from statistical information, and finally uses frequent subtree mining algorithms to excavate local rules from the subtrees of these nodes.

%% file: sections/approach.tex
\section{Approach}\label{approach}
\begin{figure}[h]
	\centering
	\includegraphics[scale=0.38]{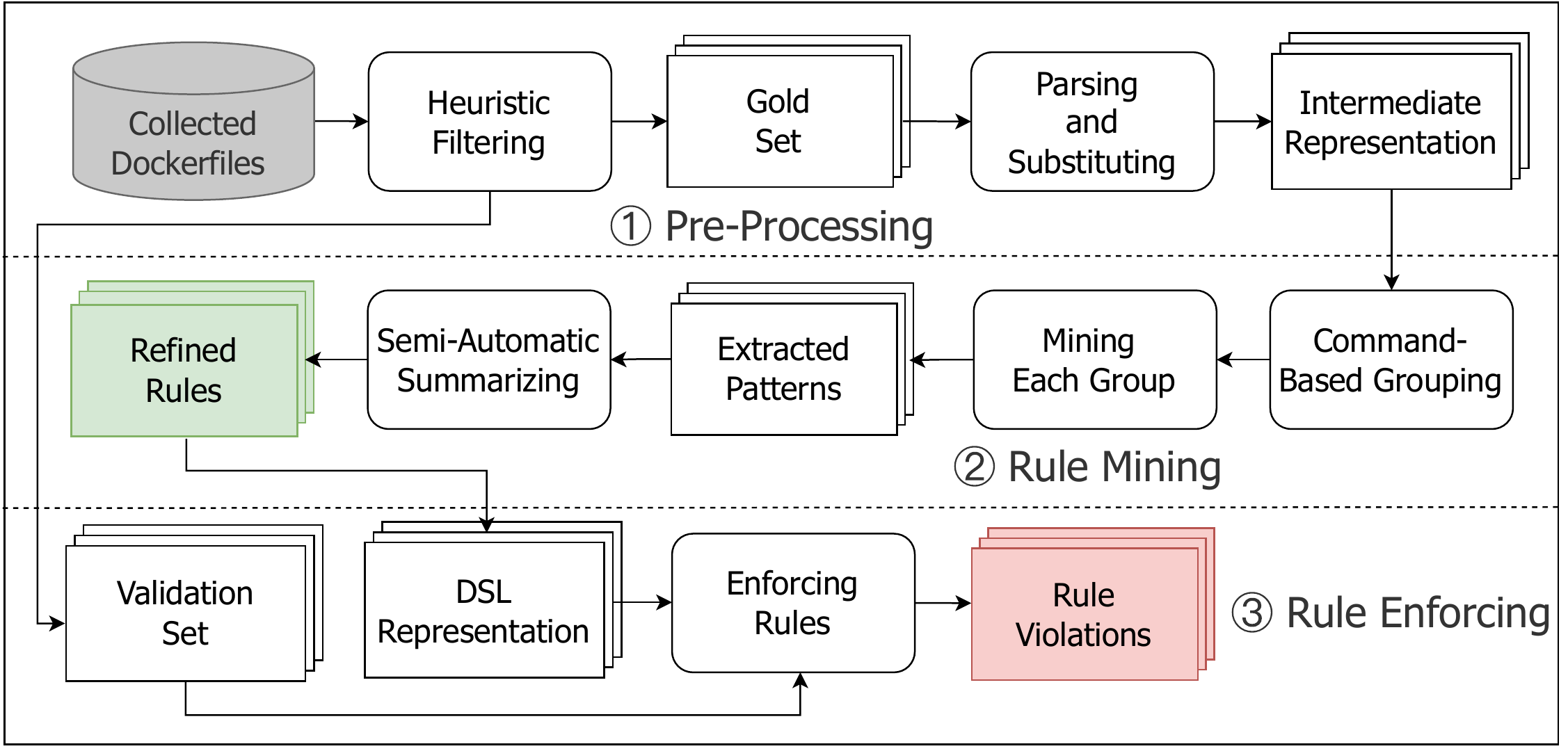}
	\caption{The workflow of  \toolname}
	\label{fig:framework}
\end{figure}

An overview of the workflow  of {\toolname} is depicted in Fig.~\ref{fig:framework}. It mainly consists of three components, i.e., pre-processing, rule mining and rule enforcing.
\begin{itemize}
	\item The pre-processing mainly involves the processing of Dockerfiles, i.e., gold set collection, file parsing and substitution,
	      and transformation of the selected Dockerfiles to an intermediate representation.
	\item {\toolname} mines the pre-processed Dockerfiles in command-based groups,  out of which preliminary patterns are extracted. To reduce the candidate size, a semi-automatic summarizing technique combining heuristic-based filtering and manual investigation is applied to generate refined rules.

	\item {\toolname} checks any input Dockerfile against the generated rules, and detects potential violations. 
\end{itemize}

\subsection{Pre-processing}

\subsubsection{Data collection}\label{sect:data}
In this step, we construct a 
gold set based on which patterns can be mined.
%
Previous studies have provided some datasets of Dockerfiles
~\cite{henkel2020, henkel2021}. However, we observe that
(1) the 
size of the dataset is relatively small (e.g., Henkel's dataset contains only about 400 Dockerfiles\footnote{Note that approximate 5,000 additional Dockerfiles were collected  as a complement, but were not used~\cite{henkel2020}.}); (2) the  dataset is 
primarily from the official Docker organization. 
As a result, although the Dockerfiles in these datasets are of high-quality,
they are under-sampled and may not be representative. 
This motivates us to re-collect dataset and construct a larger and more representative dataset of high-quality Dockerfiles 
from diversified  sources. 


To this end, we use GitHub REST APIs (mainly \code{search code}\footnote{\url{https://docs.github.com/en/rest/search\#search-code}} and \code{search repositories}\footnote{\url{https://docs.github.com/en/rest/search\#search-repositories}}) to query the entire repositories which contain Dockerfiles from GitHub (as of May 2022).  
Due to the volume of available repositories,  we 
use the number of stars  to select the most representative ones. Note that stars are often used by researchers to select GitHub projects in software engineering~\cite{borges2018}, and empirical studies have confirmed the positive correlation between popularity and quality~\cite{papamichail2016}. Particularly, we choose 1,000 stars as the threshold for the initial filtering,  resulting in 4,393 repositories and 14,260 Dockerfiles. However, high star number does not necessarily guarantee high quality. These files still suffer from various quality problems (such as syntax and size issues) for which we apply the following filtering heuristics.


\begin{itemize}
	\item We first use \emph{Hadolint} to parse  the initial set of Dockerfiles and delete those that failed the syntax checking. We also examine the size of remaining Dockerfiles, and delete those which are too small (i.e., less than 4 lines and without \code{RUN} command) and thus obviously not helpful to identify patterns.

	\item To encourage diversity, we adopt a stratified sampling strategy and sort the remaining Dockerfiles in descending order of stars per language, giving rise to five quintiles. We select the first quintile (top 20\%) of each language group for manual inspection. To this end, we hire three Docker experts, all of whom have at least 5 years of Docker-related development experience, to examine the selected Dockerfiles following the official best practice guidelines and the rules reported from existing work (e.g.,~\cite{henkel2020}). The experts adopt the majority-vote mechanism to make decisions and resolve possible conflicts. We add those files, which have passed the manual checking, to the gold set afterwards.
 
	\item Manual inspection is very time-consuming and laborious. To accelerate the process, we also record the author and affiliation information of Dockerfiles populated in the second step. To expand the gold set efficiently, we assume that Dockerfiles authored by the same developers and organizations have better quality. This assumption is based on the empirical findings that in the open source software context,  developers do not perform differently in terms of the code quality across different projects, and the developers who have more stars tend to introduce less issues~\cite{lu2018}. In this approach, we select Dockerfiles from the second quintile (20\%-40\%) of the collected dataset, and add them to the gold set.

\end{itemize}

\begin{table}
	\caption{Language distribution statistics}\label{tab:lang}
	\centering
	\begin{tabular}{lcc}
		\toprule
		Language   & Initial Set & Gold Set  \\ \midrule
		Go         & 4,361       & 372 (350) \\
		Python     & 2,765       & 354 (322) \\
		Java       & 1,374       & 214 (195) \\
		JavaScript & 1,343       & 198 (186) \\
		Shell      & 842         & 148 (122) \\
		Typescript & 831         & 113 (99)  \\
		C          & 719         & 103 (103) \\
		C++        & 710         & 131 (120) \\
		Rust       & 594         & 64  (56)  \\
		Php        & 421         & 35  (32)  \\
		Ruby       & 300         & 29  (29)  \\
		\midrule
		Total      & 14,260      & 1,761     \\
		\bottomrule
	\end{tabular}
\end{table}

After the heuristic-based filtering, we obtain a gold set $G$ of 1,761 Dockerfiles, the  distribution of  which in different programming languages is shown in Table~\ref{tab:lang}, where the number in brackets denotes those selected by the manual inspection. Note that the programming language classification of Dockerfile projects is based on the tags of the GitHub repositories. The remaining Dockerfiles which were not selected to $G$ are used as the validation set 
for experiments. (Note that the deleted Dockerfiles are not included.)


\subsubsection{Parsing and Substitution of Dockerfiles} \label{sect:parse}

\begin{figure*}[ht]
	\centering
	\includegraphics[scale=0.46]{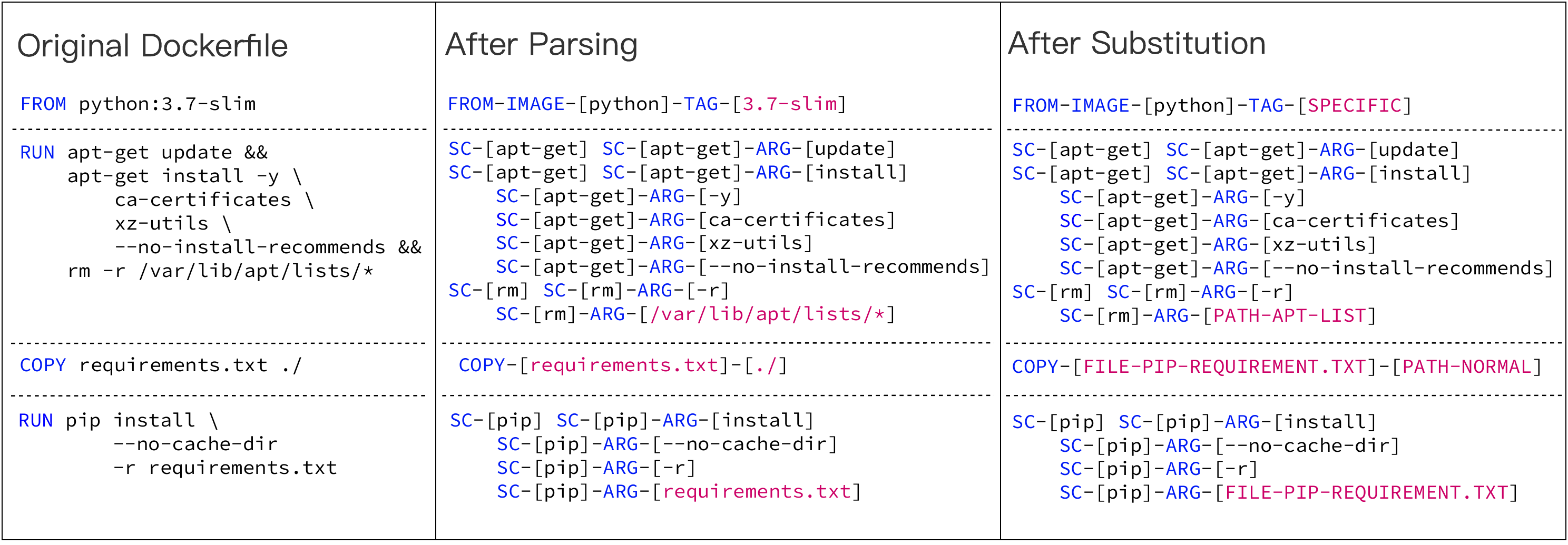}
	\caption{Before/After Parsing and Substitution\label{fig:abstract}}
\end{figure*}

We propose a parsing method to transform a Dockerfile to an intermediate representation that is convenient for the follow-up mining.
Because we pay more attention to the rules related to ``actions" rather than regular declarations, we delete declaration related instructions (those with e.g., \code{LABEL} and \code{MAINTAINER}). 

Concretely, we adopt three-phase parsing to analyze the two types of commands in Dockerfiles, i.e., Docker-specific commands and Shell scripts. The first phase is to analyze the Docker-specific commands, and the second is to parse the Shell scripts (i.e., those led by the \code{RUN} command). Finally, since there are various user-defined variables in a typical Dockerfile (e.g., file paths/names, URLs, etc.) which are too specific to be useful for the pattern mining, we abstract them away and substitute with more general, pre-defined tokens.

\begin{table}
	\caption{rules of variable substitution}\label{tab:substitution}
	\centering
	\begin{tabular}[c]{lll}
		\toprule
		Variable Regex         & Type  & Substitution             \\
		\midrule
		http://                & URL   & URL-PROTOCOL-HTTP        \\

		https://               & URL   & URL-PROTOCOL-HTTPS       \\

		ftp://                 & URL   & URL-PROTOCOL-FTP         \\

		git://                 & URL   & URL-PROTOCOL-GIT         \\

		.git                   & URL   & URL-PROTOCOL-GIT         \\

		(\textbackslash w+):// & URL   & URL-PROTOCOL-[PROTOACAL] \\
		\midrule
		var/cache/yum          & PATH  & PATH-VAR-CACHE-YUM       \\

		var/cache/             & PATH  & PATH-VAR-CACHE           \\

		var/lib/apt/lists      & PATH  & PATH-APT-LIST            \\

		src                    & PATH  & PATH-SRC-DIR             \\

		cache                  & PATH  & PATH-DOT-CACHE           \\

		\textasciitilde        & PATH  & PATH-NORMAL              \\

		.                      & PATH  & PATH-NORMAL              \\

		other path             & PATH  & PATH-NORMAL              \\
		\midrule
		.gem                   & FILE  & FILE-GEM                 \\

		.asc                   & FILE  & FILE-ASC                 \\

		.tar.gz                & FILE  & FILE-TAR-GZ              \\

		.tar.bz2               & FILE  & FILE-TAR-BZ2             \\

		.tar                   & FILE  & FILE-TAR                 \\

		.zip                   & FILE  & FILE-ZIP                 \\

		.jar                   & FILE  & FILE-JAVA-JAR            \\

		.sh                    & FILE  & FILE-SHELL-SCRIPT        \\

		.crt                   & FILE  & FILE-TLS-CERT            \\

		.pem                   & FILE  & FILE-TLS-CERT            \\

		.key                   & FILE  & FILE-KEY                 \\

		go.sum                 & FILE  & FILE-GO-SUM              \\

		go.mod                 & FILE  & FILE-GO-MOD              \\

		Cargo.toml             & FILE  & FILE-Rust-CARGO-TOME     \\

		yarn.lock              & FILE  & FILE-YARN-YARN.LOCK      \\

		package.json           & FILE  & FILE-NPM-PACKAGE.JSON    \\

		CMakeLists.txt         & FILE  & FILE-CMAKEFILEM          \\

		requirements.txt       & FILE  & FILE-PIP-REQUIREMENT.TXT \\
		\midrule

		(t$\mid$T)rue          & Other & TRUE                     \\

		(f$\mid$F)alse         & Other & FALSE                    \\

		*                      & Other & GLOB-STAR                \\
		\bottomrule
	\end{tabular}
\end{table}
In the first phase, we resort to \emph{buildkit} APIs\footnote{https://github.com/moby/buildkit/blob/master/frontend/dockerfile/parser} to parse a Dockerfile to an abstract syntax tree (AST).  By visiting each node of the tree, we can extract the command and corresponding parameter values. To distinguish with the Shell scripts, we substitute these commands with specific  annotations. As an example illustrated in Fig.~\ref{fig:abstract}, a typical \code{FROM} expression can be defined as
\begin{center}
	\code{FROM [--platform=\textless platform\textgreater ]  \textless image\textgreater [:\textless tag\textgreater ] [AS \textless name\textgreater ]}
\end{center}
It is annotated as
\begin{center}
	\code{FROM-IMAGE-[python]-TAG-[3.7-slim]}
\end{center}
in this phase as shown in the second column of Fig.~\ref{fig:abstract}.

In the second phase, we parse the Shell script led by the \code{RUN} command. To better analyze the meaning of each command in Shell, we develop a dedicated tool based on mvdan/sh\footnote{\url{https://github.com/mvdan/sh}} which is also included in the replication package. Similarly, we need to construct ASTs of the Shell script, so as to facilitate the following command extraction.  It is a non-trivial task, since we need to confirm whether each token functions as a command or as a parameter.
The Shell scripts will be parsed into different types of statements, such as \emph{assign}, \emph{call}, etc.
We mitigate this problem by analyzing ASTs.
The \emph{call} statement is made up of a command and corresponding arguments. By checking each \emph{call} statement, we can figure out the whether a token is a command or an argument.
For example, in the script \code{apt-get install unzip}, ``\code{unzip}" is used as a parameter of ``\code{apt-get}", but the ``\code{unzip}" in \code{unzip example.zip} is a command. 
We also annotate the other parsed Shell commands in a similar way as illustrated in Fig.~\ref{fig:abstract}.

In the third phase, we transform the parsed Dockerfile to an intermediate representation. The main purpose is to abstract away the unnecessary details which are not useful for pattern mining. 
We observe that there are common variables and files in the Dockerfiles which need to be unified to adapt to the association rule mining algorithm, so we propose substitution rules for variables which can be summarized as 35 heuristic substitution rules and are classified into four different categories, i.e., URLs, file paths, file names, and others, given in Table~\ref{tab:substitution}. For example, we can extract the image name as a key information from the \code{FROM} instruction, keep the image name, and substitute the image tag with either \code{LATEST} or \code{SPECIFIC} depending on the corresponding values in Dockerfiles. 

For URLs in Dockerfile, we focus on the protocols and file types. For example, ``\code{https://abc.com/\\a/download.zip}" is transformed to ``\code{URL-PROTOCOL-HTTPS}" and ``\code{FILE-ZIP}" sequentially. Similarly, we abstract local paths. For example, in Fig.~\ref{fig:abstract} we substitute the second argument of the \code{COPY} command by ``\code{PATH-NORMAL}", while the  path of ``\code{/var/lib/apt/lists/*}" in the \code{RUN} instruction is converted to ``\code{PATH-APT-LIST}".  We also abstract file names in Dockerifles. For example,  \code{requirements.txt} in Fig.~\ref{fig:abstract} is transformed to ``\code{FILE-PIP-REQUIREMENT.TXT}". We ignore the path prefixes of such files. Namely, ``\code{/app/requirement.txt}", ``\code{requirement.txt}", ``\code{pip-requirements\\.txt}" and other similar local files will be replaced with a unified intermediate representation of ``\code{FILE-PIP-REQUIREMENT.TXT}".

As shown in Fig.~\ref{fig:abstract}, in the final intermediate representation, we use ``\code{SC-[\$cmd]}" to mark Shell command, ``\code{SC-[\$cmd]-ARG-[\$arg]}" to mark parameters of the corresponding commands. The irrelevant symbols, such as ``\code{\&\&}" and ``\code{\textbackslash}",  will be deleted. For example, in Fig~\ref{fig:abstract}, \code{pip install --nocache-dir -r requirements.txt} is parsed into a sequence ``\code{SC-[pip], SC-[pip]-ARG-[inst\\all], SC-[pip]-ARG-[--no-cache-dir], SC-[pip]-ARG-[FILE-PIP-REQUIREMENT.TXT]}".

Note that the representation is important for the later pattern/rule mining as we want the rules to be as general as possible and not to overfit to specific details.

\subsection{Rule Mining}

After pre-processing, we are now in a position to mine patterns from the Dockerfiles. Notably, we employ frequent sequence mining algorithms to identify frequent patterns. The underlying observation is that in high-quality Dockerfiles, sequences that conform to specific rules are more likely to occur. We focus on sequential pattern mining because Dockerfiles are largely sequential (no branch or loop in Docker-specific commands).

Various data mining methods are available for discovering frequent patterns, which can be basically classified into three categories, i.e, itemset-based mining, sequence-based mining, and tree-based mining. We choose the frequent sequence mining algorithm since the alternatives may suffer from effectiveness and/or efficiency issues.  On one hand, Dockerfiles consist of instructions that are structured sequentially and contain nested shell statements. Itemset-based frequent sequence mining algorithms ignore such order information, producing a large amount of redundant results, which requires additional manual efforts. In addition, the ordering information is not preserved in the returned results, which introduces further difficulty to the follow-up rule construction. On the other hand, tree-based frequent subtree mining algorithms  usually perform well in processing source code with control flow information. However, Dockerfiles do not have conditional or loop control flows, and these structures rarely appear in nested shell statements. Therefore, frequent subtree mining algorithms do not show advantages and may increase the complexity of the mining process instead, resulting in efficiency issues.


\subsubsection{Command based grouping}
Shell-related commands take a majority of the collected Dockerfiles. To better identify the patterns among these commands, we divide the gold set into multiple groups based on the Shell commands. All the Dockerfiles in the gold set containing same Shell command will be put in a group denoted by that command. Analysing the intermediate representations, we find 77 commands annotated by ``\code{SC}'' which denotes the Shell command type. It is common that one Dockerfile contains multiple Shell commands, such as the example in Fig.~\ref{fig:abstract}, we adopt a replicated grouping strategy. In other words, the Dockerfile with multiple Shell commands will be replicated and included in all these corresponding command groups.



\subsubsection{Mining}
We employ the PrefixSpan algorithm (cf.\ Section~\ref{sect:spm}) to extract patterns in each command group derived from the previous step. 
However, the mined frequent subsequences 
are way too many. To reduce the size, we select the \emph{maximal} sequential patterns from the output of PrefixSpan. Maximal sequential patterns is defined as
those where no sequence is a subsequence of that sequence. For example, ``\code{pip install-r requirements.txt}" is a subsequence of ``\code{pip install --no-cache-dir -r requirements.txt}", so the latter is a maximal sequential pattern. 
Since we treat subsequences with the support value greater than a given threshold equally, selecting maximal sequential patterns to represent other patterns can effectively reduce the data size without losing information.

\subsubsection{Semi-automatic summarization}


Based on the maximal sequence patterns discovered in each command group, we can refine and extract the corresponding rules. It is difficult to be fully automated to obtain these rules, since domain expertise is required to decide whether or not they are indeed implicit rules for Dockerfiles. Therefore, we incorporate human participation in this step. Despite the preliminary reduction of candidate set in the previous step via maximal subsequence, the remaining candidate set is still very large. To further boost productivity, we again use heuristics to prune irrelevant ones in each group.

\begin{figure}[h]
	\centering
	\includegraphics[scale=0.55]{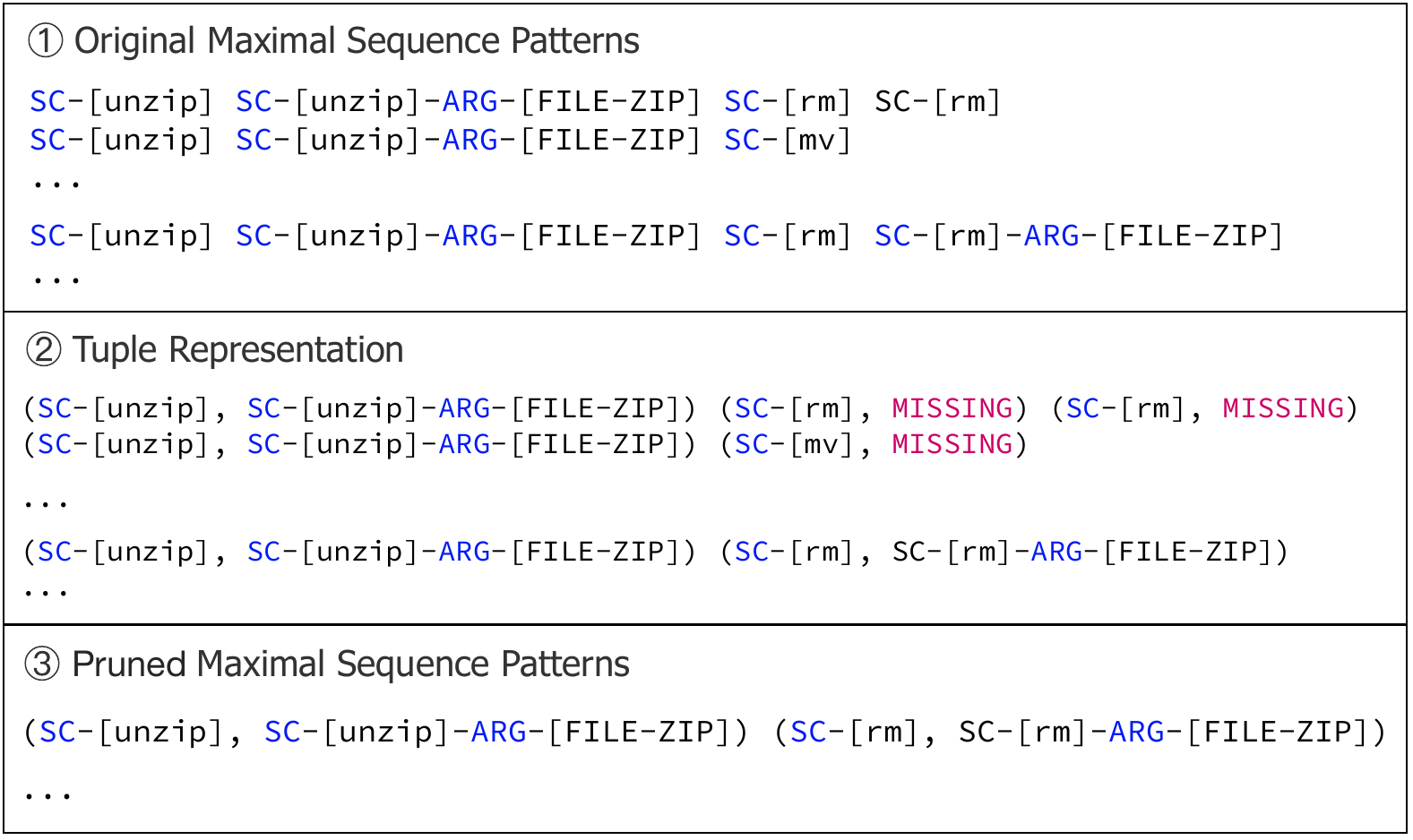}
	\caption{\label{fig:tuple}Pruning of Tuple-Represented Patterns}
	\label{fig:framewok}
\end{figure}

Given the maximal sequence patterns obtained in each command group, we use tuples to represent them. Each tuple has two parts, i.e., \emph{command}, and \emph{parameters}. The former denotes the specific command, and the latter denotes the corresponding parameters. As an example shown in Fig.~\ref{fig:tuple}, the pattern excerpt in the first part is selected from the \code{unzip} command group, and the second part is the tuple representation. Since sequence mining just considers co-occurrences of items, it is high likely that there are incomplete tuples in the returned patterns. We assume that the patterns with incomplete tuple information is less likely to be a potential rule. The underlying rationale is: 1) if the command part is missing, the parameters alone do not make sense to be included; 2) if the parameter part is missing, it means there is no frequent co-occurrence of the command and any of its parameters above the threshold support value. Then we can consider that the probability of extracting rules from this pattern is much lower than those of complete patterns. As an example in Fig.~\ref{fig:tuple}, we can observe that the parameters of ``\code{SC-[rm]}" and ``\code{SC-[mv]}" are missing.  Therefore, we can prune the patterns containing these incomplete tuples to further reduce the size. 


As a result, the number of the maximal sequences left in each collection could be greatly reduced.  We can then manually select and summarize the corresponding rules from each collection. For example, in the last pattern of the third part of Fig.~\ref{fig:tuple}, we can summarize an unzip-related rule:
\begin{quote}
	when a compressed file is decompressed by \code{unzip}, the original compressed file should be deleted to save space.
\end{quote}

Following the classification of rules give by~\cite{henkel2020}, we summarize two types of rules for  the remaining patterns of each command group, i.e., syntax, and semantic. Syntax rules refer to those regarding the grammatical regulations of command usage. For instance, there should be two parameters of command \code{CP}; semantic rules describe those regarding the operational meanings of the commands, as shown by the \code{unzip} example in Fig.~\ref{fig:tuple}.


\subsection{Rule Enforcement}

As mentioned before, we have identified two categories of implicit rules, i.e., syntax rules and semantic one. For the former type, violation detection can be conducted through a common Shell linter (e.g., ShellCheck\footnote{\url{https://github.com/koalaman/shellcheck}}), so we mainly focus on the  semantic rule violation detection. Based on the relation of the elements within the rule, we classify  the semantic rules into four types as follows.
\begin{itemize}
	\item  $P \Rightarrow Q$, which means that when $P$ appears, there must be $Q$ after it, otherwise there is a violation.
	\item $(P_1|P_2|\cdots |P_n) \Rightarrow (Q_1|Q_2|\cdots |Q_n)$, which means that when any one of $P_1, \cdots, P_n$ appears, there must be one of  $Q_1, \cdots, Q_n$ after it, otherwise there is a violation.
	\item $P \Leftarrow Q \Rightarrow R$, which means that when $Q$ appears, there must be $P$ before it, and there must be $R$ after it, otherwise there is a violation.
	\item $SPECIAL$, which denotes special rules to be enforced separately.
\end{itemize}

\begin{figure}[h]
	\centering
	\includegraphics[scale=0.72]{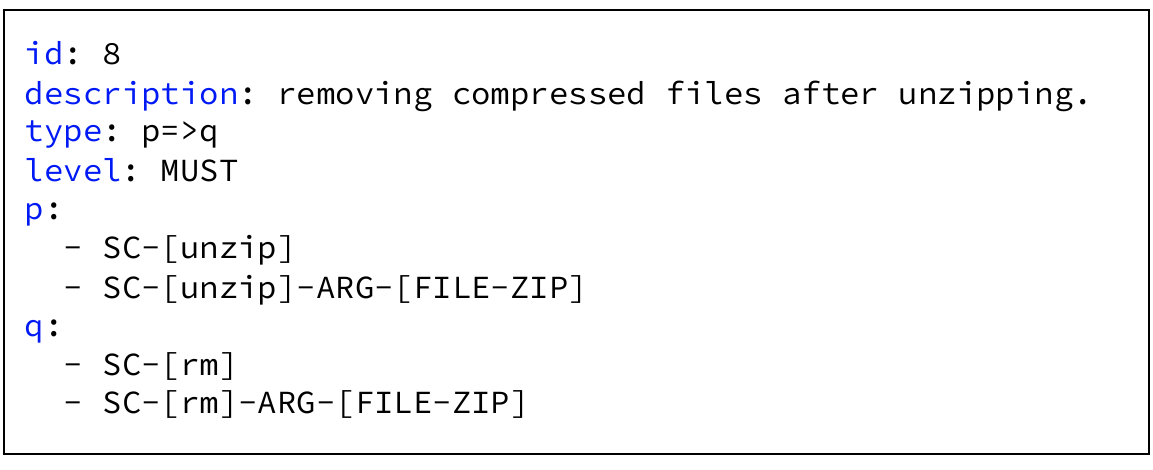}
	\caption{Illustration of YAML-Based DSL Rule Description}
	\label{fig:DSL}
\end{figure}

To facilitate rule interpretation and the follow-up violation detection, we
use a custom YAML-based domain specific language (DSL) to describe each semantic rule based on the above classification. Fig.~\ref{fig:DSL} shows an example DSL for the \emph{unzip} rules. All the rules will be encoded in such DSL style and used as a configuration file to drive the following detection process.


\begin{algorithm}
	\small
	\caption{Detection Algorithm} \label{alg:Detection}
	\KwData{Dockerfile $D$ and Rule list $R$}
	\KwResult{List of Violation $V$}
	$D'$ = ParseAndSubstitution($D$)

	\ForEach{$r$ in $R$}{
	\uIf{$r.type$ is $P \Rightarrow Q$}{
		\SetKwProg{Def}{def}{:}{}
		\Def{check(seq)}{

			$p$ $\leftarrow$  list of positions where $r.P$ last appeared in $seq$ \;

			\uIf{p is None} {
				\Return{True} \;
			}

			$q$ $\leftarrow$  boolean value of whether $r.Q$ appears in $seq[p_{max}+1:]$ \;

			\uIf{q is False} {
				\Return{False} \;
			}

			\Return{check(seq[0:$p_{min}$])} \;

		}
		\uIf{check($D'$) is False} {
			$V.add$($r$) \;
		}

	}
	\uElseIf{$r.type$ is $(P_1|...|P_n) \Rightarrow (Q_1|...|Q_n)$ }{

	$p$ $\leftarrow$  list of positions where $r.(P_1|...|P_n)$ last appeared in $D'$ \;

	\uIf{p is None} {
		Continue next loop \;
	}


	$q$ $\leftarrow$  boolean value of whether $r.(Q_1|...|Q_n)$ appears in $D'[p_{max}+1:]$ \;

	\uIf{q is False} {
		$V.add$($r$) \;
	}

	}
	\uElseIf{$r.type$ is $P \Leftarrow Q \Rightarrow R$}{
	%
	%

	\While{True}{
	$q$ $\leftarrow$  list of positions where $r.Q$ first appeared in $D'$ \;
	\uIf{q is None}{
		Break loop \;
	}
	$p$ $\leftarrow$  boolean value of whether $r.P$ appears in $D'[0:q_{min}]$ \;
	$r$ $\leftarrow$  boolean value of whether $r.R$ appears in $D'[q_{max}+1:]$\;
	\uIf{p and r is not True}{
		$V.add$($r$) ;
	}
	$D'$ $\leftarrow$ $D'[q_{max}+1:]$
	}

	}
	\uElseIf{$r.type$ is SPECIAL}{
		Execute the process that belongs to the specific rule;
	}
	}
\end{algorithm}

Our detection process is shown in Algorithm 1. The algorithm requires two inputs, viz., the rules in the DSL format and the Dockerfile to be detected. We firstly parse the Dockerfile as described in Section~\ref{sect:parse} and obtain the processed $D'$ (Line 1). Then we iterate each rule and process it based on its type.  If the rule is of the form $P \Rightarrow Q$, we locate the last position where $P$ occurs in $D'$, and split $D'$ from this position into two parts, i.e., $D'_L$ (the left part) and $D'_R$ (the right part). If $Q$ does not appear in $D'_R$, it is regarded as a violation of this rule (Line 4-10); otherwise, repeat  the above process on $D'_L$ until $P$ cannot be found (Line 11).

If the rule is of the form $(P_1|P_2|\cdots |P_n) \Rightarrow (Q_1|Q_2|\cdots|Q_n)$, we locate the position where any one of $P_1, \cdots, P_n$ lastly appears in $D'$, and similarly split $D'$ from this position into two parts, i.e., $D'_L$ and $D'_R$. If none of $Q_1, \cdots, Q_n$ appear in $D'_R$, it is regarded as a violation of this rule (Line 15-20).

If the rule is of the form $P \Leftarrow Q \Rightarrow R$, we need to find all the positions where $Q$ occurs. For each position, we split $D'$ from this position and obtain $D'_L$ and $D'_R$.  If $P$ and $R$ cannot be found in $D'_L$ and $D'_R$ respectively, it is considered as a violation of this rule (Line 22-31).

Finally, if the rule is of the SPECIAL type, the processing flow that belongs to the rule is executed to determine whether there is a violation of that rule (Line 33). The processing of this type of rules must be tailored on the case-by-case basis.
	For example, for Rule 32 in Table~\ref{tab:rules}, we leverage the Shell parser to check whether  the embedded Shell commands following the \emph{RUN} instruction in the Dockerfile contain ``set -eux''.

%% file: sections/exp.tex
\section{Evaluation}\label{exp}
In this section, we conduct experiments to evaluate our approach. Particularly, we aim to answer the following research questions (RQs).

\begin{description}
	\item[RQ1.] How effective is the rule mining component of {\toolname}?
	\item[RQ2.] How effective is the violation detection component of {\toolname}?
	\item[RQ3.] How efficient is the overall {\toolname} approach?
\end{description}

\subsection{Experimental setup}
The experiments were conducted on a server with an Intel Xeon 2.3GHz 32-core CPU and 32GB RAM running Arch Linux. The prototype is implemented with \emph{Go} v1.18 and \emph{Python} v.3.10.4.
The baselines compared in the experiments are \emph{Hadolint}~\cite{hadolint} and \emph{Binnacle}~\cite{henkel2020}. 
As mentioned in Section~\ref{approach}, we collect Dockerfiles from GitHub, based on which the datasets are populated. Particularly, there are mainly three datasets used in our experiments.

\begin{description}
	\item[D1.] The initial dataset collected from GitHub consists of 14,260 Dockerfiles, but with duplicates. We
		remove the duplicates, resulting in a dataset with 12,066 Dockerfiles.
	\item[D2.] This is the Gold set of Dockerfiles we construct from D1 (cf.\ Section~\ref{sect:data}). It contains 1,761 Dockerfiles.
	\item[D3.] This is another Gold Set provided by \emph{Binnacle}, which includes 405 Dockerfiles.
\end{description}

For the hyperparameter, {\toolname} only needs to set 
the support value threshold for subsequence frequency. The threshold value directly affects the number of output pattern candidates, mining efficiency and manual inspection effort. As common in machine learning, we tune this hyper-parameter via experiments and find that 40\% is an appropriate value.

\subsection{RQ1: The effectiveness of rule mining}
\begin{figure}
	\centering
	\includegraphics[scale=0.33]{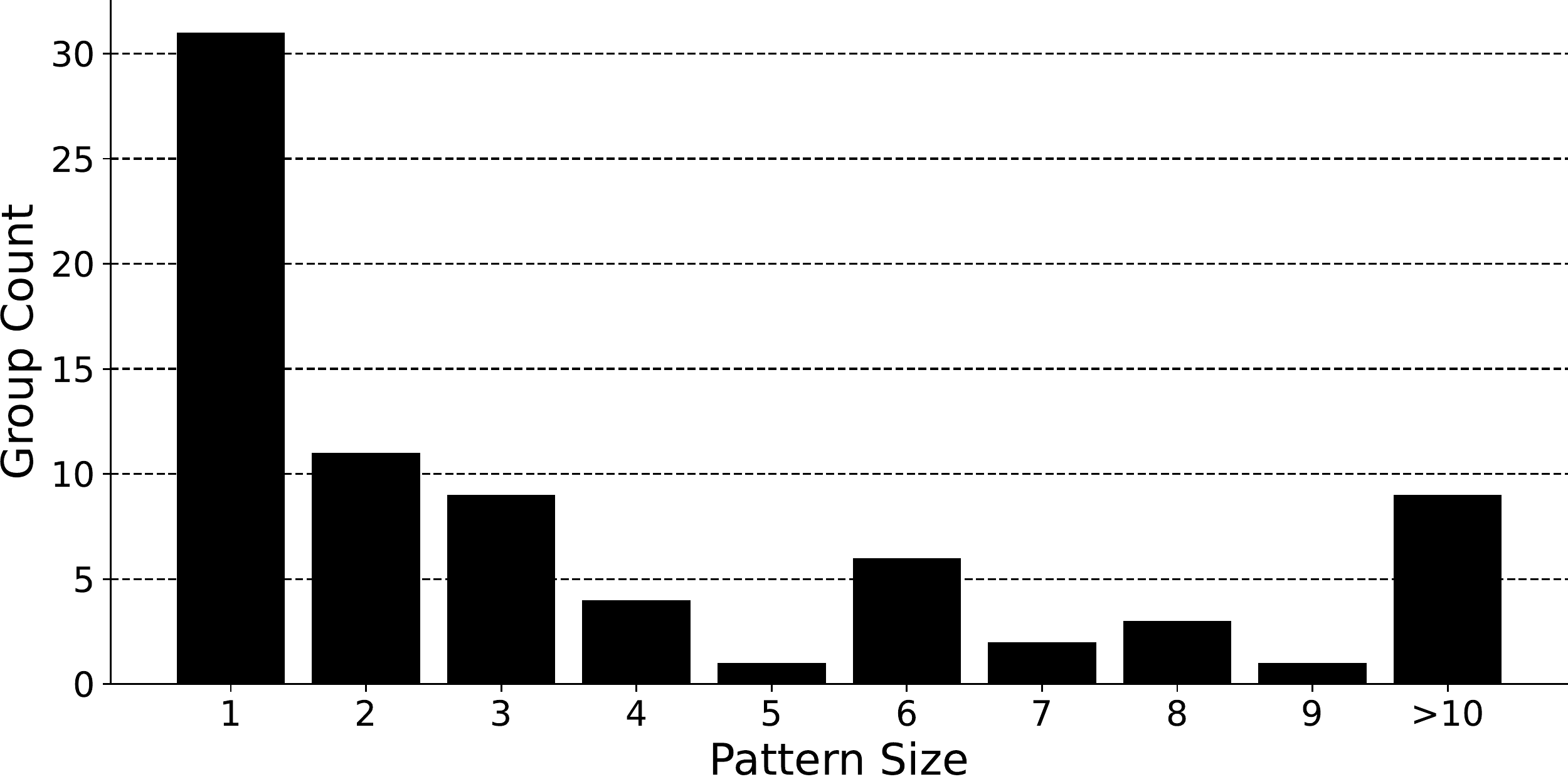}
	\caption{Distribution of rules in groups}\label{fig:pattern_count}
\end{figure}

\begin{table}
	\caption{Syntactic rules mined by {\toolname}\label{tab:syntax_rule}}
	\centering
	\begin{tabular}[c]{llcl}
		\toprule
		Id & Rule                 & Id & Rule         \\
		\midrule
		1  & go build             & 11 & mvn package  \\
		2  & go get               & 12 & gem install  \\
		3  & bundle install       & 13 & make install \\
		4  & npm install -g       & 14 & cargo build  \\
		5  & tar -C               & 15 & mv PATH PATH \\
		6  & ssh-keygen -t        & 16 & cat PATH     \\
		7  & sh -s                & 17 & ls PATH      \\
		8  & yarn build           & 18 & cp PATH PATH \\
		9  & addgroup/groupadd -g & 19 & touch PATH   \\
		10 & git clone            &    &              \\

		\bottomrule
	\end{tabular}
\end{table}

\begin{table*}
	\caption{Semantic rules mined by {\toolname}\label{tab:rules}}
	\centering
	\footnotesize
	\begin{tabular}[c]{llllcc}
		\toprule
		Id & Rule Description                                         & Rule Type                                 & Level & Confidence & Lift \\
		\midrule
		1  & apk add using arg. --no-cache                            & $P \Rightarrow Q$                         & M     & 86\%       & 4.43 \\
		2  & pip install using arg. --no-cache-dir                    & $P \Rightarrow Q$                         & M     & 55\%       & 1.68 \\
		3  & pip install using requirement.txt                        & $P \Rightarrow Q$                         & E     & 66\%       & 3.48 \\
		4  & curl using arg. -f                                       & $P \Rightarrow Q$                         & E     & 77\%       & 1.39 \\
		5  & curl with url type https                                 & $P \Rightarrow Q$                         & M     & 89\%       & 1.58 \\
		6  & wget with url type https                                 & $P \Rightarrow Q$                         & M     & 82\%       & 1.49 \\
		\rowcolor{xgray}
		7  & git clone with url type https                            & $P \Rightarrow Q$                         & E     & 96\%       & 1.72 \\
		\rowcolor{xgray}
		8  & removing compressed files after unzipping                & $P \Rightarrow Q$                         & M     & 70\%       & 1.51 \\
		9  & tar something then remove                                & $P \Rightarrow Q$                         & M     & 64\%       & 1.43 \\
		10 & gpg using arg. --batch                                   & $P \Rightarrow Q$                         & E     & 45\%       & 9.31 \\
		11 & gpg using arg. --keyserver                               & $P \Rightarrow Q$                         & E     & 45\%       & 9.31 \\
		12 & gpg using .asc file then remove the .asc file            & $P \Rightarrow Q$                         & E     & 60\%       & 9.12 \\
		13 & dnf install using arg. -y                                & $P \Rightarrow Q$                         & M     & 76\%       & 1.57 \\
		14 & mkdir using arg. -p                                      & $P \Rightarrow Q$                         & E     & 61\%       & 1.02 \\
		\rowcolor{xgray}
		15 & chown using arg. -r                                      & $P \Rightarrow Q$                         & E     & 61\%       & 0.89 \\
		16 & rm using arg. -rf                                        & $P \Rightarrow Q$                         & E     & 77\%       & 1.63 \\
		17 & yum install using  arg.  -y                              & $P \Rightarrow Q$                         & M     & 84\%       & 1.78 \\
		18 & zypper install using arg. -y                             & $P \Rightarrow Q$                         & M     & 81\%       & 1.72 \\
		19 & apt-get install using arg. -y                            & $P \Rightarrow Q$                         & M     & 72\%       & 1.53 \\
		20 & apt-get install using arg. --no-install-recommends       & $P \Rightarrow Q$                         & M     & 77\%       & 1.63 \\
		21 & configure using arg. -build                              & $P \Rightarrow Q$                         & M     & 85\%       & 7.83 \\
		\midrule
		22 & apt-get update prefix apt-get install                    & $P \Leftarrow Q \Rightarrow R$            & M     & 76\%       & 2.09 \\
		\rowcolor{xgray}
		23 & go build using multi-stage                               & $P \Leftarrow Q \Rightarrow R$            & E     & 91\%       & 4.47 \\
		\rowcolor{xgray}
		24 & java build using multi-stage                             & $P \Leftarrow Q \Rightarrow R$            & E     & 72\%       & 6.67 \\
		\midrule
		\rowcolor{xgray}
		25 & clean cache after using conda to install packages        & $(P_1|...|P_n) \Rightarrow (Q_1|...|Q_n)$ & M     & 72\%       & 7.21 \\
		26 & clean cache after using apt-get/dpkg to install packages & $(P_1|...|P_n) \Rightarrow (Q_1|...|Q_n)$ & M     & 68\%       & 2.81 \\
		27 & clean cache after using zypper to install packages       & $(P_1|...|P_n) \Rightarrow (Q_1|...|Q_n)$ & M     & 75\%       & 8.82 \\
		28 & clean cache after using dnf to install packages          & $(P_1|...|P_n) \Rightarrow (Q_1|...|Q_n)$ & M     & 61\%       & 9.77 \\
		29 & clean cache after using yum/rpm to install packages      & $(P_1|...|P_n) \Rightarrow (Q_1|...|Q_n)$ & M     & 71\%       & 5.73 \\
		\rowcolor{xgray}
		30 & using sha to verify the downloaded file                  & $(P_1|...|P_n) \Rightarrow (Q_1|...|Q_n)$ & E     & 56\%       & 1.54 \\
		\rowcolor{xgray}
		31 & using gpg to verify the downloaded file                  & $(P_1|...|P_n) \Rightarrow (Q_1|...|Q_n)$ & E     & 42\%       & 1.32 \\
		\midrule
		\rowcolor{xgray}
		32 & set -eux to print command and quick fail in shell script & $Special$                                 & E     & N/A        & N/A  \\
		33 & using useradd to avoid last user to be root              & $Special$                                 & E     & N/A        & N/A  \\
		34 & using groupadd/addgroup to avoid last user to be root    & $Special$                                 & E     & N/A        & N/A  \\
		\bottomrule
	\end{tabular}
\end{table*}

As mentioned above, D2 is the Gold dataset based on which we apply {\toolname}. In the first step, we group the parsed Dockerfiles based on the commands and obtain 77 groups in our case. Then we mine the frequent patterns from each group. The average size of the preliminary output patterns of each group is 4,515. However, we are only interested in maximal subsequence patterns, which reduces the size to 18, approximately 0.4\% of the original size. Though the size is greatly reduced, it is still too large for manual examination. After pruning the patterns with incomplete tuple information, the average pattern size in each group is further reduced to 4, shrunk by 77.8\%. The distribution of pattern size among the command groups is shown in Fig.~\ref{fig:pattern_count}.

We then ask the three Docker experts to examine the resultant patterns of each group. Finally we obtain 53 rules, including 34 semantic ones and 19 syntactic ones. Generally, semantic rules are more interesting and useful in practice, since syntactic rules are easier to be detected through conventional Linter tools. Table~\ref{tab:syntax_rule} and Table~\ref{tab:rules} show these syntactic and  semantic rules respectively.  Particularly, we assign two levels to the semantic rules, i.e., ``\code{MANDATORY}" and ``\code{ENCOURAGED}".  The former level means that the rule should be followed rigorously, while the latter means that they are strongly suggested. The details of the level information, as well as the confidence ratio and lift ratio~\cite{tuffery2011} of the discovered patterns, are summarized in Table~\ref{tab:rules}.


We find that these rules cover all the 15 filtered rules identified in the \emph{Binnacle} toolset, and 19 of them are
actually included in the total 23 rules manually summarized in \emph{Binnacle}. 10 rules match those summarized in
\emph{Hadolint}. Therefore, in total, among the 34 semantic rules identified by {\toolname}, 24 of them match those 
manual rules devised by previous work (4 rules exist in both \emph{Hadolint} and \emph{Binnacle}). Interestingly, we also find 
9 rules by {\toolname}, which are highlighted in Table~\ref{tab:rules}, were not identified before by \emph{Binnacle} or 
\emph{Hadolint}. 

We observe that a considerable amount of rules are related to Shell commands. Interestingly, these rules are not general, since some can only be applied in the context of Dockerfiles. Representative examples include rules 25-29 listed in Table~4, i.e., ``deleting the cache generated during package installation using a package management tool''. In a typical independent Shell environment, keeping these caches is beneficial, because reusing them can save bandwidth or speed up future installation tasks. However, when building a Dockerfile, these caches will not be used again. So including these caches will inevitably result in  an unnecessarily large built image. Such examples show that the existing general-purpose Shell best practices should not be simply taken for granted. 

The 9 new rules we find are all semantic ones, and may have negative consequence if they are violated. For instance, Rule 8 states that, when building a docker image, if the original file is not deleted after decompressing the file, a large amount of storage space will be wasted, because it makes no sense to keep the original file in the docker image. This corresponds to the illustrative example in Fig.~\ref{exam2b}. Rules 23 and 24 suggest that, if programs written in static languages (such as Go and Java) need to be compiled when building Docker images, it is supposed to use the multi-stage build strategy so as to avoid generating intermediate files during compilation. Interestingly, this rule is confirmed by a question ``How to reduce my java/gradle docker image size?'' posted in  the StackOverflow websit.\footnote{\url{https://stackoverflow.com/questions/40958062/how-to-reduce-my-java-gradle-docker-image-size}} where the developer complained that the final image size was high up to 1.1 GB due to all the unnecessary files included. The accepted answer pointed that using multi-stage build can keep the jar files only and get rid of those unnecessary intermediate files. 

\begin{table}
	\caption{Rules mined by \emph{Binnacle} and {\toolname } in different datasets\label{tab:rule_count} }
	\centering
	\begin{tabular}[c]{ccc|cc}
		\toprule
		\multirow{2}*{Dataset} & \multicolumn{2}{c}{\toolname} & \multicolumn{2}{c}{\emph{Binnacle}}                     \\
		\cmidrule{2-3}\cmidrule{4-5}
		                       & Semantic                      & Syntactic                    & Semantic & Syntax \\
		\midrule
		D3                     & 7                             & 11                           & 4        & 12     \\
		\midrule
		D2                     & 33                            & 19                           & 7        & 17     \\
		\bottomrule
	\end{tabular}
\end{table}

Since {\toolname} and \emph{Binnacle} operate on different Gold sets to mine patterns, it might be unfair to conclude that {\toolname} is more effective. To give a fair comparison, we also run the two tools on the same Gold sets. i.e., \emph{D2} and \emph{D3}. The comparative results are given in Table \ref{tab:rule_count}. 

We observe no substantial difference in mining syntactic rules. 
This is because all the syntax-related rules seem to be local and can be mined  by either the frequent subtree mining algorithm used by \emph{Binnacle}, or the frequent sub-sequences mining algorithm used by \toolname.
However, {\toolname} shows advantages in mining semantic association rules. This is because, after replacing variables, we retain the semantics of the text and can mine the relationship among commands. As a result, we can conclude that {\toolname} is more effective to identify implicit semantic rules in Dockerfiles.



\subsection{RQ2: The effectiveness of violation detection in {\toolname}}

To assess the rule violation detection component of {\toolname}, we compare with \emph{Hadolint}, since \emph{Binnacle} does not provide such a functionality.  We generate the test dataset by sampling the initial validation dataset D1. Namely, we randomly select 300 Dockerfiles in the initially collected dataset excluding the Dockerfiles from the Gold set.


\begin{table}[H]
	\caption{Rules covered in Hadolint}\label{tab:rule_same_hadolint}
	\centering
	\begin{tabular}[c]{ll}
		\toprule
		Id & Rule Description                                         \\
		\toprule
		1  & apk add using arg. --no-cache                            \\
		2  & pip install using arg. --no-cache-dir                    \\
		13 & dnf install using arg. -y                                \\
		17 & yum install using arg. -y                                \\
		18 & zypper install using arg. -y                             \\
		19 & apt-get install using arg. -y                            \\
		20 & apt-get install using arg. --no-install-recommends       \\
		26 & clean cache after using apt-get/dpkg to install packages \\
		28 & clean cache after using dnf to install packages          \\
		29 & clean cache after using yum/rpm to install packages      \\
		\bottomrule
	\end{tabular}
\end{table}

%

\begin{table}[H]
	\caption{Various metrics in violation detection\label{tab:rule_enforcement_metric}}
	\centering
	\begin{tabular}[c]{lccccccc}
		\toprule
		Approach & TP & FP & TN & FN  & Precision & Recall & F-measure \\
		\midrule
		\toolname & 195 & 19 & 85 & 1 & 0.911 & 0.995 & 0.951      \\
		\emph{Hadolint} & 182 & 0 & 104 & 14 & 1.0 & 0.929 & 0.963     \\
		\bottomrule
	\end{tabular}
\end{table}


Among the identified semantic rules reported by {\toolname} shown in Table~\ref{tab:rules}, only 10 (given in Table~\ref{tab:rule_same_hadolint}) are also reported by \emph{Hadolint}. To ensure fairness, we only consider these rules in comparing violation detection capabilities 
on the test set.
In our experiment, {\toolname} and \emph{Hadolint} report 215 and 182 Dockerfiles with rule-violations, respectively. To further investigate the result 
we again ask the three Docker experts to manually annotate each file in the dataset for the violations of the 10 rules. 
We then calculate the Precision, Recall, and F-measure for each tool. The result is shown in Table~\ref{tab:rule_enforcement_metric}.

It can be observed that {\toolname} and \emph{Hadolint} 
demonstrate different advantages in terms of Precision and Recall. {\toolname} reports almost no false negatives (FN), meanwhile \emph{Hadolint} reports no false positives (FP). Their F-measure are almost at the same level. \emph{Hadolint} is slightly higher than {\toolname} (96.3\% versus 95.1\%). 
Upon examining the detection results, we find that this is caused by the internal logic of their detection methods.  {\toolname} uses a sequence-based method to detect the violation of rules. Some rare but valid sequences may cause a false negative reported by our approach. For example, in Rule 19, in most cases, developers  write the \code{-y} parameter following \code{apt-get install} command. However, \code{apt-get -y install} is also a valid expression which is semantically equivalent, but will be falsely identified as a violation by {\toolname}. \emph{Hadolint}, on the other hand, uses a Shell parser in detection and can accurately identify such a case. However, the downside is the report of false negatives. 
For example, in a real-world Dockerfile\footnote{\url{https://github.com/siaorg/sia-task/blob/f0bb2c4fd40b752bbd571e17232db7c24ad041c4/sia-task-docker/scheduler-docker/scheduler/Dockerfile\#L12}}, when \code{RUN yum clean all \&\& yum makecache \&\& yum install ...} appears, \emph{Hadolint} finds that the same \code{RUN} statement contains both software installation and cache clearance actions, so it verdicts that this case does not violate the rule which causes a false negative (corresponding to Rule 19). {\toolname}, on the other hand, can accurately detect the violation based on the sequence information whether there is a cache clearance action after the last installation action in the RUN statement. In addition, we investigated the only false negative case in {\toolname}. The false report was caused by a Dockerfile\footnote{\url{https://github.com/bitpay/bitcore/blob/88318365e65509a386376f39cd6b4579063cf654/.docker/rippled.Dockerfile}} that violated Rule 19 exactly at the last use of the \code{apt-get install} command. In this command, the \code{-y} parameter was missing and there was another command with the \code{-y} parameter after it. 



\subsection{RQ3: The efficiency of the overall approach}

In this research question, we mainly consider the performance, particularly of the running time. As a comparison, we run \emph{Binnacle} and {\toolname} on all the three  datasets mentioned above, and collect their running time in the parsing and rule mining phases, respectively. The details of  time cost comparison are given in Table \ref{tab:time_cost}.

\begin{table}[H]
	\caption{Time Cost of \toolname\ and \emph{Binnacle}}\label{tab:time_cost}
	\centering
	\begin{tabular}[c]{lcc|cc}
		\toprule
		\multirow{3}*{Dataset} & \multicolumn{2}{c}{\toolname} & \multicolumn{2}{c}{\emph{Binnacle}}                                           \\
		\cmidrule{2-3}\cmidrule{4-5}
		                       & \multirow{2}*{Parsing (s)}    & Rule                                & \multirow{2}*{Parsing (s)} & Rule       \\
		                       &                               & mining (s)                          &                            & mining (s) \\
		\midrule
		D3 (405)               & 3                             & 201                                 & 62                         & 1,028      \\
		\midrule
		D2 (1,761)             & 14                            & 257                                 & 264                        & 1,386      \\
		\midrule
		D1 (12,066)            & 68                            & 1,134                               & 337                        & N/A        \\
		\bottomrule
	\end{tabular}
\end{table}

We also collect the time spent by {\toolname} and \emph{Hadolint} on rule detection in Table~\ref{tab:time_cost_detect}.
As shown in the table, our algorithm is very fast in rule detection and is more than twice as fast as that of \emph{Hadolint}.

\begin{table}[H]
	\caption{Time Cost in Voilation Detection of \toolname\ and \emph{Hadolint}}\label{tab:time_cost_detect}
	\centering
	\begin{tabular}[c]{lc|c}
		\toprule
		\multirow{2}*{Dataset} & \toolname          & \emph{Hadolint}    \\
		                       & Rule Detection (s) & Rule Detection (s) \\
		\midrule
		D3 (405)               & 4                  & 8                  \\
		\midrule
		D2 (1,761        )             & 15                 & 40                 \\
		\midrule
		D1 (12,066)            & 70                 & 345                \\
		\bottomrule
	\end{tabular}
\end{table}

It can be observed that the efficiency of DRIVE is higher than \emph{Binnacle} in the data preprocessing part and rule mining part. In data preprocessing, our processor can efficiently process Dockerfile into sequence form. While in the rule mining part, on the one hand, because the sequence mining algorithm we chose has the advantage in speed, but also because our mining method is designed to be parallel, the mining work between each group is independent of each other and can run in parallel. Therefore, the running time of the tool can be significantly accelerated.

Based on the above analysis, we can conclude that \toolname\ can extract Dockerfile rules and detect violations of a large volume of Dockerfiles very efficiently.

%% file: sections/discuss.tex
\section{Discussion}\label{discuss}
In Section~\ref{sect:data}, we collect the initial dataset containing the Dockerfiles deemed to have a high quality, but the actual mining process indicates that this is not the case. As shown in RQ2, 45\% of Dockerfiles have at least one rule violation. For example, 2,976 Dockerfiles use the \code{pip} command. However, only 607 of them use \code{pip} with  \code{--no-cache-dir} argument. We believe that, when writing Dockerfile, using \code{pip}  with \code{--no-cache-dir} is a rule that should be followed. This rule does not have any side effects in the Dockerfile context, but the benefits are apparent. 

As mentioned before, our approach is sequence based. This has certain advantages, for instance, it is easy to mine and can be extended to new, emerging Dockerfiles. Moreover, to be compatible with the mining process, the obtained rules are also specified as properties of sequences (cf.\ Table~\ref{tab:rules}), which are easier to understand comparing to the previous work which specify the rules based on ASTs. However, a slight disadvantage  is that our violation detection may not be as precise as the approach using ASTs. In RQ2, we observe a (albeit only marginally) higher false positive. It is possible to convert sequence-based rules to AST-based rules, but it may require more human involvement, which is against the philosophy of the current work. We leave as future work how to combine these two approaches in a better way. 

In this paper, we assume that Dockerfiles are largely sequential (no branch or loop in either Docker-specific commands or Shell scripts). However, in some rare cases, there exist branch statements or loop statements in the Shell scripts of Dockerfile's \code{RUN} instruction. Though the commands in such statements could be successfully parsed, the execution sequence does not match the assumption of sequential pattern mining. Therefore, in our experiment, we remove the Dockerfiles with such statements in the Gold set to reduce the potential noises. 
We also notice that sometimes developers  move the Shell commands following \code{RUN} to a separate script file such as \code{install.sh}, in this case, we did not analyze the contents of the separated Shell scripts as well, and these files are also excluded from the Gold set.

We focus on Dockerfiles for two reasons. Firstly, Docker is the \emph{de facto} industry standard in the container ecosystem. Secondly, Dockerfiles specify the building instructions which directly determine the resultant image quality. Moreover, Dockerfiles can also be reused by some other Docker-compatible container tools such as Podman\footnote{\url{https://podman.io/}} and Buildah\footnote{\url{https://buildah.io/}}. Therefore, our approach could be used as-is to improve the quality of the built image of those containers. 
More generally, in DevOps, configuration files can be basically categorized into imperative and declarative styles. For imperative configuration files that incorporate a significant amount of sequential information, such as Dockerfile and Chef\footnote{\url{https://www.chef.io/}} configuration files, DRIVE would perform well given an abundant of golden data with moderate adaptation if necessary. On the other hand, for purely declarative configuration files such as those used in Kubernetes\footnote{\url{https://kubernetes.io/}} and Puppet\footnote{\url{https://www.puppet.com/}}, where the sequential information is typically irrelevant, replacing the sequence mining algorithm in DRIVE with a frequent itemset mining algorithm could yield better results.



\subsection{Threats to Validity}\label{validity}
\paragraph{Construct validity} This aspect of validity is related with the degree to which variables represent the concepts~\cite{sjoberg2022}. In our approach, we solely rely on the sequence patterns mined from the Dockerfiles to extract potential rules. To balance the efficiency and accuracy, we leverage a set of heuristics and abstraction rules to accelerate the mining process. These heuristics are based on domain experts and observations. To mitigate the potential errors introduced in this process, we double check these heuristics and hire human experts to manually check the selected samples after processing. For the detection part of {\toolname}, we use the typical metrics such as precision, recall, and F-measure to evaluate the performance, which are widely used in the literature.
\paragraph{Internal validity} The internal threats  are mainly introduced from the bias of the collected data. To mitigate this, we collect an initial Dockerfile dataset from a diversity of domains and programming languages. We select projects with high star numbers as the initial data source. This metric is a direct indicator for popularity~\cite{borges2016} and widely used as a criterion to select GitHub projects in empirical studies in software engineering~\cite{borges2018}. Popular projects usually attract more attention and more participation which are crucial for open-source software quality assurance. Therefore, projects with more stars are more likely to be of higher quality. In our experiments, we set the threshold to be 1,000 stars. In the remaining projects, we use a set of heuristics, including both tool support and human examination, to further select high quality Dockerfiles. The selection criteria are based on the public tags and commonly used in similar research practice.
\paragraph{External validity} The external validity is  mainly about the generalization issues of the proposed approach. We admit that  it is impractical to collect all the high quality Dockerfiles based on which all rules could be automatically extracted. However, we demonstrated that with the current collected data, our approach could already find many interesting rules, some of which have not been covered in the state-of-the-art tools. On the other hand, the methodology of our approach, i.e., applying data mining techniques to other software artifacts to find potential patterns, has also well been demonstrated by related work in the literature~\cite{bian2018nar,li2005pr,liang2016antminer}.

%% file: sections/related.tex
\section{Related Work}\label{related}
In this section, we briefly review three threads of relevant work in literature, i.e. empirical studies on Dockerfiles, automatic Dockerfile analysis, and pattern extraction from software artifacts. 

\noindent \textbf{Empirical studies on Dockerfiles.} Cito \emph{et al.} performed the first empirical study on open-source ecosystem of Docker~\cite{cito2017}. Particularly, they investigated the quality and evolution behaviors of Dockerfiles. A considerable proportion of Dockerfiles suffered from various quality issues, calling for effective quality check integration. Wu \emph{et al.} observed that Dockerfile smells are common in the collected Docker projects, taking up to 84\% of the population, and there existed co-occurrence between certain types of Dockerfile smells~\cite{wu_characterizing_2020}. Eng and Hindle analysed the change history of a large-scale Dockerfiles and reconfirmed the many code smells reported by previous studies with a slightly decreasing trend over the recent years~\cite{eng2021}. Zerouali \emph{et al.} examined the Debian-based Docker images over a three-year period, and found more than $90\%$ of \emph{community} images did not use the \emph{apt upgrade} command in the building process, leading to potential outdated packages in the generated image~\cite{zerouali2021multi}. Ksontini \emph{et al.} investigated the refactoring history of 68 projects and identified a set of technical debt issues with inappropriate Dockerfiles, such as build time, image size, maintainability~\cite{ksontini21}. Interestingly, a more recent empirical study 
~\cite{azuma_empirical_2022} revealed that a specific type of technical debt, i.e., self-admitted technical debt takes up to 3.4\% in the explored datasets via manual investigation over the comments.

Despite the different focuses of these empirical studies, they confirm the necessity of quality check for Dockerfiles.

\smallskip
\noindent\textbf{Automatic Dockerfile analysis.} Perhaps the closest work to ours is \emph{binnacle} 
~\cite{henkel2020} where  Henkel \emph{et al.} propose a phased parsing approach to analyse Dockerfiles, based on which Docker specific commands and Shell commands could be modeled as ASTs. Then the tree association rules (TARs) could be obtained via frequent sub-tree mining afterwards. However, the work is susceptible to high computation cost and could only identify intra-directive rules under the same local node. Differently, our work treats the commands sequentially, which gives a performance advantage, and inter-directive rules could be identified.

DockerMock
~\cite{li2021dockermock} aims to timely detect Dockerfile faults before actual building. It mocks the execution of Dockerfile instructions based on the parsed ASTs within fuzzy contexts. Similarly, \emph{shipwright}
~\cite{henkel2021} also attempts to repair the broken Dockerfiles to pass the building requirements through static analysis. Some other work has been proposed to address the duplicates or type-2 clone issues among multiple Dockerfiles~\cite{oumaziz2019handling,tsuru2021type}. Different from our work, the emphasis of such work is mainly to detect faults or duplicates instead of best practice violations. 

DockerizeMe attempts to automatically infer the dependencies of Python code snippets and generate Dockerfiles to deliver the environment configuration~\cite{horton2019dockerizeme}. Meanwhile, RUDSEA proposed by Hassan \emph{et al.} can generate Dockerfile changes as updates along with fast software evolution by analysing changes of software environment asumptions and their impacts~\cite{hassan_rudsea_2018}. Such line of work mainly focuses Dockerfile synthesis instead of pattern mining as in our approach. Xu \emph{et al.} described a specific kind of Dockerfile smells, termed as ``Temporary File Smell", which denotes the unnecessary temporary files are shipped in the final built Docker images. They propose dynamic analysis and static analysis approaches to detect and fix such smells in Dockerfiles~\cite{xu_dockerfile_2019,lu2019empirical}. However, in our work we adopt a data-driven way to identify patterns in general and  detect violations of such rules correspondingly.

\smallskip
\noindent\textbf{Pattern extraction from software artifacts.} Li and Zhou proposed a frequent itemset mining based approach,  i.e., PR-Miner to extract implicit, undocumented programming rules from large software codebase. Thousands of rules could be extracted within less than 1 minute. The tool can also be leveraged to detect the violations of the extracted rules~\cite{li2005pr}.  Sun \emph{et al.} extended typical static analysis tools with dependence-based rule mining technique, and more project-specific programming rules could thus be discovered~\cite{sun2012extending}. Liang \emph{et al.}~\cite{liang2016antminer} applied a frequent itemset mining algorithm, i.e., \emph{FPClose}~\cite{grahne2003efficiently} to the pre-processed codebase by program slicing. With the extracted rules, the approach can effectively  detect a number of subtle bugs
that have been missed previously. Their subsequent work, NAR-miner, employed a similar technique, but to extract negative association rules from large-scale codebase, and detect their violations to find bugs~\cite{bian2018nar}. Cao \emph{et al.} adopted a learning-to-rank approach to mine specification rules in Java programs by combining 38 measures~\cite{cao2018rule}.



Besides mining conventional programs, some approaches work on Shell scripts. Dong \emph{et al.} \cite{dong2023bash} presented a large-scale empirical study of Bash usage based on  over one million open-source scripts found in GitHub repositories, identifying frequently used language features and common smells in these scripts. D'Antoni \emph{et al.} \cite{D'Antoni2017NoFAQ} presented NoFAQ, a tool that suggests possible fixes for commonly occurring errors in command-line tools by using a set of rules expressed in a domain specific language and evaluated the tool on 92 benchmark problems through a crowd-sourcing interface. Mazurak \emph{et al.} \cite{Mazurak2007ABASH} presented ABASH, a tool for statically analyzing Bash scripts that can detect certain common program errors leading to security vulnerabilities.
They reported experiments with 49 bash scripts, identifying 20 as containing bugs of varying severity while yielding only a reasonable number of spurious warnings. 
Different from our work, these approaches do not consider Docker environment, and  thus the patterns found may not be adequate in the Docker context as discussed previously.

Apart from mining codebase,  other kinds of software artifacts could also be mined to extract interesting patterns, for example, error patterns from software revision history~\cite{livshits2005dynamine}, past-time temporal rules from execution traces~\cite{lo2008mining}, specification rules from configuration files~\cite{santolucito2017}. 
These approaches deal with different types of software artifacts than ours. 